\documentclass[prl,aps,twocolumn,showpacs]{revtex4}
\usepackage[dvips]{graphicx}
\usepackage{dcolumn}
\newcommand{\be}{\begin{equation}}
\newcommand{\ee}{\end{equation}}
\newcommand{\bea}{\begin{eqnarray}}
\newcommand{\eea}{\end{eqnarray}}
\newcommand{\nn}{ \nonumber}
\newcommand{\ds}{\displaystyle}
\begin{document}
\topmargin=-20mm
%    \Large

\title{ Local Features of the Fermi Surface Curvature and the Anomalous Skin Effect in Metals}

\author{Natalya A. Zimbovskaya}

\affiliation
{Department of Physics and Electronics, University of Puerto Rico at Humacao, Humacao, PR 00791}

\begin{abstract}
 In this paper we present a theoretical analysis of the effect of local geometrical
 structure of the Fermi surface on the surface impedance of a metal at the anomalous
 skin effect. We show that when the Fermi surface includes nearly cylindrical and/or
 flattened segments it may significantly change both magnitude and frequency dependence
 of the surface impedance. Being observed in experiments these unusual frequency
 dependencies could bring additional information concerning fine geometrical
 features of the Fermi surfaces of metals.
   \end{abstract}

\pacs{ 73.21 Cd; 73.40. --C}
%\vspace{2mm}
\date{\today}
\maketitle

\section{I. Introduction}

 It is well known that electromagnetic waves incident at the
surface of a metal cannot penetrate deeply inside. Actually, the
field inside the metal vanishes at the distances of the order of $
\delta $ from the surface. This effect is called the skin effect,
and the characteristic depth $ \delta $ is called the skin depth.
The suppression of the electromagnetic field inside the metal
originates from the response of conduction electrons, and it occurs
when the frequency $ \omega $ of the incident wave is smaller than
the electrons plasma frequency $ \omega_p. $ The latter is the
characteristic frequency for the response of the conduction
electrons system to an external disturbance. When $ \omega >
\omega_p $ the electrons are too slow to respond, and the
electromagnetic field penetrates into the metal without decay. Due
to the skin effect the incident electromagnetic field could affect
condition electrons only when they move inside the layer of the
thickness $ \delta $ near the metal surface. The skin depth depends
on the electric conductivity of the metal $ \sigma $ and on the
frequency $ \omega $ of the incident wave as well. Increase in $
\sigma $ and/or $ \omega $ leads to the decrease in the skin depth.
At high frequencies $ \tau^{-1} \ll \omega \ll \omega_p\ (\tau $ is
the scattering time for conduction electrons) and low temperatures,
$ \delta $ may become smaller than the electrons mean free path $ l.
$ When the condition $ \delta < l $ is satisfied the effect is
referred to as the anomalous skin effect. At the anomalous skin
effect the response of a metal to an incident electromagnetic wave
is determined with the electrons moving in the skin layer nearly in
parallel with the surface of the metal sample. These ``efficient"
electrons are associated with a few small ``effective segments" on
the Fermi surface (FS).  The remaining electrons stay in the skin
layer only for a very short while which prevents them from
responding to the electromagnetic field.

A theory of the anomalous skin effect in metals was first proposed
more than five decades ago by A. B. Pippard \cite{1} and G. E.
Reuter and E. H. Zondheimer \cite{2}, and R. B. Dingle \cite{3}
using an isotropic model for a metal. The main results of these
studies were presented in some books where high frequency phenomena
in metals were discussed \cite{4}. Then the theory was further
developed to make it applicable to realistic metals with anisotropic
Fermi surfaces \cite{5,6,7,8,9}. It became clear that the response
of conduction electrons to an external electromagnetic field under
the anomalous skin effect depends on the Fermi surface (FS)
geometry, especially its Gaussian curvature
 $\ K({\bf p}) = 1/R_1 ({\bf p}) R_2 ({\bf p}), \ $
 where $ R_{1,2}\bf (p) $ are the principal radii of curvature.
For the most of real metals FSs are complex in shape, and their
curvature turns zero at some points. These points could be partitioned
in two classes. First, these exist zero curvature points where only one
of the principal radii has a singularity, whereas another one remains
finite. Usually, such points are combined in lines of zero curvature.
The latter are either inflection lines or they label positions of
nearly cylindrical strips on the FSs. Also, some points could be
found where both principal radii tend to infinity. These points are
set out separately, and the FSs are flattened in their vicinities.

When a FS includes points of zero curvature it leads to an enhancement of
the contribution from the neighborhoods of these points to the electron
density of states (DOS) on the FS. Normally, this enhanced contribution is
small compared to the main term of the DOS which originates from the major
part of the FS. Therefore it cannot produce noticeable changes in the
response of the metal when all segments of the FS contribute  essentially
equally. However, when the curvature turns zero at some points on an
``effective" part of the FS, it can give a sensible enhancement in the
number of efficient electrons and, in consequence, a pronounced change
in the response of the metal to the disturbance.

It has been shown that when the FS includes nearly cylindrical
and/or flattened segments, noticeable changes may be observed in the
frequency and temperature dependencies of sound dispersion and
absorption \cite{10,11,12,13}. Also, the shape and amplitude of
quantum oscillations in various characteristics of a metal could be
affected by  of the FS local geometry in the vicinities of the
extremal cross sections. Qualitative anomalies in the de Haas-van
Alphen oscillations associated with cylindrical pieces of the FSs
were considered in Refs. \cite{14,15}. Similar anomalies in quantum
oscillations in the static elastic constants and the velocity of
sound were analyzed in \cite{16,17}.

Here, we concentrate on the analysis of possible manifestations of
the FS local geometry in the surface impedance of a metal at the
anomalous skin effect. In this case the main contribution to the
surface impedance of a metal originates from electrons moving nearly
in parallel with the surface of the metal. These electrons are
efficient quasiparticles, and they belong to the ``effective" part
of the FS. The effects of the FS geometry on the metal response at
the anomalous skin effect  were analyzed before \cite{9,18} adopting
some simplified models for the FS. The purpose of the present work
is to carry out a general analysis whose results are independent on
particularities in energy-momentum relations and could be applied to
a broad class of metals.

\section{II. results and discussion}

We consider a metal filling the half-space $ z < 0 $. A plane
electromagnetic wave is incident on the metal surface making a right
angle with the latter.  To analyze the response of the metal to the
wave we calculate the surface impedance:
   \be % f1
  Z_{\alpha\beta} = E_\alpha (0) \bigg/ \int_0^{-\infty} J_\beta (z)
  d z
  \ee
   Here, $ \alpha,\beta = x,y;\ E_\alpha(z) $ and $ J_\beta (z) $
   are the components of the electric field $ \bf E $ and electric
   current density $ \bf J, $ respectively.
 Considering the anomalous skin effect we can limit our analysis to the case
 of specular reflection of electrons from the surface. Then the surface
 impedance tensor has the form:
  \be % f2,3
Z_{\alpha \beta} = \frac{8 i\omega}{c^2} \int_0^\infty
\bigg(\frac{4 \pi i \omega}{c^2} \sigma - q^2 I\bigg )_{\alpha \beta}^{-1} d q.
 \ee
 Here, $ \omega $ and $ \bf q $ are the frequency and the wave vector of the
 incident wave, respectively $ ({\bf q} = (0,0,q)); \ \sigma $ is the electron
 conductivity tensor, and $ I_{\alpha \beta} = \delta_{\alpha \beta}. $

To proceed we assume that the FS has a mirror symmetry in a momentum space
relative to a plane $ p_z = 0.$ To simplify calculations of the electron
conductivity we divide each sheet of the FS in segments in such a way that
the momentum $ \bf p $ is a one-to-one function of the electron velocity
$ \bf v$ over a segment. The segments may coincide with the FS sheets.
Also, it could happen that  some sheets include a few segments. This
depends on the FS shape. In calculation of the conductivity we carry out
integration over each segment using spherical coordinates in the velocity
space, namely, the velocity magnitude at the $j$-th segment $ v_j, $ and
the spherical angles $\theta, \varphi. $ So, the element of the surface
area is given by the expression:
$ d A_j = \sin \theta d \theta d \varphi/|K_j(\theta,\varphi)| $
 where $ K_j (\theta,\varphi)$ is the Gaussian curvature of the $j$-th FS
 segment. Summing up contributions from all these segments we obtain:
    \bea %f3,4
&&\sigma_{\alpha\beta} (\omega, q)= \frac{i e^2}{4 \pi^3 \hbar^3 q}
 \sum_j \int d \varphi
  \nn \\ &&
\times
\int \frac{ n_\alpha n_\beta   \sin \theta d \theta}
{\big |K_j(\theta, \varphi)\big | \big [(\omega + i/\tau) /q v_j - \cos
\theta \big ]}.
          \eea
   Here, $ n_{\alpha,\beta} = v_{j\alpha,\beta}/v_j, $ and $ \tau $ is the
   electron scattering time. The limits in the integrals over $ \theta, \varphi $
   are determined with the shape of the segments. We remark, however, that the
   effective strips on the FS are determined by the condition $ v_z \approx 0 $ for
   efficient electrons move in parallel with the metal surface at $ z=0. $
   Therefore, the upper limit in the integral over $ \theta $ in the terms
   corresponding to the segments including the effective strips must equal $ \pi/2. $
   In the following calculations we omit the term $ i/\tau $ in the Eq. (3)
   assuming $ \omega \tau \gg 1$ which is typical for the anomalous skin
   effect in good metals. Using Eq. (3) we can easily write out the
   expressions for the conductivity components. We have:
  \bea %f4,3,4
&& \sigma_{xx} (\omega, q)= \frac{i e^2}{4 \pi^3 \hbar^3 q}
\sum_j \int d \varphi
   \nn \\ &&\times
\int \frac{d \theta \cos^2 \varphi \sin^3 \theta}
{\big |K_j(\theta, \varphi)\big | \big [(\omega + i/\tau) /q v_j - \cos
\theta \big ]}.
          \eea
  Another conductivity component $ \sigma_{yy} $ is described with the similar
  expression where $ \cos^2 \varphi $ in the integrand numerator is replaced
  by $ \sin^2 \varphi. $ In further calculations we assume for simplicity that
  the chosen $ z $ axis coincides with a high symmetry axis for the FS, so that
  both conductivity and surface impedance tensors are diagonalized.

The main contribution to the surface impedance under the anomalous skin effect
comes from the region of large $ q $ where $ \omega/q v \ll 1 .$ To  calculate
the corresponding asymptotic expressions for the conductivity components we
expand the integrand in the Eq. (4) in powers of $ \omega/qv $.
Then we can write the well known result for the principal
term in the expansion of the conductivity component $\sigma_{xx}(\omega, q) $:
  \be %f5,4,5
 \sigma_0 (q) = \frac{e^2}{4 \pi^3 \hbar^3 q} \sum \limits_l
 \int d \varphi \frac{\cos^2 \varphi}{\big |K_l (\pi/2, \varphi) \big |}
 \equiv \frac{e^2}{4 \pi \hbar^3 q} p_0^2.
           \ee
 The same asymptotics could be obtained for $ \sigma_{yy}, $ so
the indices are omitted for simplicity here and in following
expressions. Summation over $ l $ is carried out over all segments of
the FS containing effective strips which correspond to $ \theta = \pi/2 \ (v_z = 0) $
and the curvature $ K_l (\pi/2, \varphi ) $ is supposed to take finite
and nonzero value at any point of any effective strip. For a spherical
FS $ p_0 $ equals the Fermi momentum $ p_F.$ In realistic metals the two
are not equal but have the same order of magnitude. In general, $ p_0 $
is determined by the Eq. (5). Then we can calculate the next term in the
expansion of conductivity in powers of $ \omega/qv. $ For a FS whose
curvature everywhere is finite and nonzero we arrive at the result
   \be %f6,5,6
 \sigma_1 (\omega, q) = \sigma_0 (q) \frac{i \omega}{q v_0}.
           \ee
  Here, the velocity $v_0 $ has the order of the Fermi velocity $ v_F: $
  \bea %f7
  \frac{1}{v_0} &=& \frac{2}{\pi^2p_0^2} \sum_l \int d \varphi
\int_{\alpha_l}^{\pi/2} \frac{d\theta \cos^2 \varphi \sin\varphi}{\cos^2\theta}
   \nn \\ \nn \\ &\times &
 \left[\frac{1 + \cos^2 \theta/\cos^2 \alpha_l}{|K_l (\pi/2,\varphi)|
 v_l (\pi/2,\varphi)} - \frac{\sin^2 \theta}{|K_l (\theta,\varphi)|
 v_l (\theta, \varphi)}    \right]
   \nn \\ \nn \\ &-&
 \frac{2}{\pi^2 p_0^2} \sum_{j\neq l} \int d \varphi \int_{\theta < \pi/2}
\frac{d \theta \cos^2 \varphi \sin^3 \theta}{|K_j (\theta,\varphi)|
v_j (\theta,\varphi) \cos^2 \theta}. \nn\\
   \eea
   Here, the lower limit $ \alpha_l $ in the integral over $ \theta $
   in the first term takes on values determined by the FS shape
   $ (\alpha_l < \pi/2). $ The second term corresponds to the
   contribution from the FS segments which do not include effective
   lines. For the spherical FS we have $v_0=\pi v_F/4.$

When the curvature at any effective line turns zero, it changes the
 asymptotics for the conductivity, as we show below. First, we assume
  that the curvature becomes zero at a whole effective line passing
  through one of the segments of the Fermi surface. Keeping in mind
  that $ z $ axis in the chosen reference system runs along a high  order
  symmetry axis we can present the relevant energy-momentum relation in the form:
  \be %f8,6
  E{\bf(p)} = E_1 (p_x,p_y) + E_2(p_z).
  \ee
  Near the effective line where $ v_z \equiv \partial E_2/\partial p_z = 0 $
  we can approximate $ E_2 (p_z) $ as follows:
   \be %f9,7
  E_2 (p_z) \approx E_0 \left(\frac{p_z - p^*}{p^*} \right)^{2l};\qquad l\geq 1.
  \ee
  Here, $ E_0, p^* $ have the dimensions of the energy and momentum,
  respectively; $ p_z = p^* $ corresponds to the effective line.

In the vicinity of the effective line we can write the following
expression for the curvature of the FS corresponding to the Eq. (8):
  \be %f10,8
  K{\bf (p)} = \frac{1}{v^4} \frac{\partial v_z}{\partial p_z}
\left (v_y^2 \frac{\partial v_x}{\partial p_x} +
v_x^2 \frac{\partial v_y}{\partial p_y} -
2 v_x v_y \frac{\partial v_x}{\partial p_y} \right ).
   \ee
  The value of the curvature at $ v_z = 0 $  is determined by the factor
  $\ \partial v_z/\partial p_z \sim [(p_z - p^*)/p^*]^{2l-2}. $ The curvature
  becomes zero at the effective line when $ l > 1. $ Expressing this factor as a
  function of velocity (which is necessary to carry out integration over a region
  in the velocity space) we get $ \partial v_z/\partial p_z \sim (v_z)^{-\beta} $
  where $ \beta = -1 +1/(2l -1). $

So, we can use the following approximation for the curvature
$K (\theta, \varphi) $ at $ \theta \le \pi/2: $
     \be %f11,6, 7
K(\theta,\varphi) = W (\theta,\varphi) (\cos \theta)^{-\beta},
          \ee
 In this expression, the function $ W(\theta,\varphi) $ everywhere
 assumes finite and nonzero values, and the exponent $ \beta $ takes on
 negative values which correspond to the line of zero curvature  at
 $\theta =\pi/2. $  In the close vicinity of this line the FS is nearly
 cylindrical in shape. The closer $ \beta $ to $-1,$ the closer to
 a cylinder is the effective strip on the FS. The contribution to
 the conductivity from the nearly cylindrical segment on the FS is given by:
     \bea % f7,8
\!\!\! & &
\sigma_a (\omega, q) = \frac{i e^2 \omega}{2 \pi^3 \hbar^3 q}
\Bigg [ \int d \varphi \int_\alpha^{\pi/2} \! d \theta
\Bigg (\frac{\sin^2 \theta}{\big |W(\theta, \varphi)\big |v(\theta,\varphi)}
 \nn \\ \nn \\  \! \!\!\! & - &
\frac{1}{\big |W(\pi/2, \varphi)\big
|v(\theta, \varphi)} \Bigg) \frac{\cos^2 \varphi \sin \theta
(\cos \theta)^\beta}{\big (\omega /q v (\theta, \varphi)\big)^2 - \cos^2 \theta}
\nn \\  \nn \\ \! \!\!\! &+& \!\!\!
 \int
\frac{ d \varphi \cos^2 \varphi }{\big|W(\pi/2, \varphi)
\big |v(\pi/2, \varphi)} \int_{\alpha}^{\pi/2} \!\!\!
  \frac{d \theta \sin \theta (\cos \theta)^\beta}
{\big(\omega /q v (\theta, \varphi)\big)^2 - \cos^2 \theta} \Bigg ] . \nn \\
          \eea

 Using this asymptotic expression we can calculate the ``anomalous"
 contribution to the conductivity $ \sigma_a (\omega, q) $ for small
 $ \omega/q v. $ Introducing the largest magnitude of the velocity
 on the effective line $ v_a $ we have:
    \bea %f13,8,9
 \sigma_a (\omega, q) &=& \rho \sigma_0 (q)
\left (\frac{\omega}{q v_a} \right )^\beta \left [1 - i \tan
\bigg (\frac{\pi \beta}{2} \bigg) \right ] ,
   \\ \nn \\
 \rho &\approx& \frac{1}{\pi p_0^2} \int \frac{d \varphi
 \cos^2 \varphi}{|W (\pi/2, \varphi)|},
      \eea
  Comparing Eq. (14) with the definition for $ p_0^2 $ introduced
  earlier by Eq. (5) we see that $ \rho $ is a dimensionless factor
  whose value is determined with the relative number of the ``effective"
  electrons concentrated at the nearly cylindrical effective segment.

The value of the contribution to the conductivity from the
"anomalous" effective strip depends on the character of the
curvature anomaly at given strip, and on the relative number of
effective electrons concentrated here as shown in the Fig. 1. In
this figure we display plots of $ |\sigma (\omega,q)/\sigma_0 (q)|
 \equiv |1 + \sigma_a(\omega,q)/\sigma_0 (q)| $ versus $ \omega/q v. $
 When the parameter $ \rho $ takes on values of the order or greater
 than 0.1 (the number of effective electrons associated with the
 anomalous sections on the FS is comparable to the total number
 of the effective electrons), the term $ \sigma_a (\omega,q) $
 can predominate over $ \sigma_0 (q) $ and determine the conductivity
 value at large $ q. $ This occurs when the shape parameter $ \beta $
 accepts values not too close to zero, and the curvature anomaly at
 the effective line is well pronounced. When either $ \rho $ or
 $ \beta $ or both are very small in magnitude, the main approximation
 to the conductivity is described with Eq. 5 as well as for a metal
 whose FS curvature is everywhere nonzero. Nevertheless, in such
 cases the term $ \sigma_a (\omega, q) $ also is important for it
 gives the first correction to the principal term in the expression
 for the conductivity.

 \begin{figure}[t]
\begin{center}
\includegraphics[width=8.6cm,height=4.4cm]{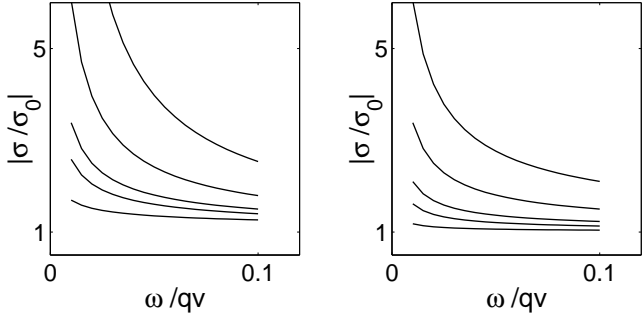}
%[width=5.0cm,height=6.6cm]{M1.eps}%{g2.eps}
\caption{
Conductivity component $ \sigma_{xx} (\omega,q ) $ including the
contribution from a zero-curvature segment on the FS at large $ q \
(\omega/qv \ll 1). $ Left panel: the curves are plotted at
$ \rho = 0.2 $ and $ \beta = -0.8,-0.7, -0.6,-0.4,-0.2 $
from the top to the bottom. Right panel: the curves are
plotted at $ \beta=- 0.6 $ and $ \rho = 0.2, 0.1,0.05,0.03, 0.01 $
from the top to the bottom.
} \label{rateI}
\end{center}
\end{figure}

Also, the anomalous contribution to the conductivity could appear
when the FS is flattened at some points belonging to an effective
segment. To avoid lengthy calculations we illustrate the effect of
such points on the conductivity using a simple expression
representing the energy momentum relation near the point of
flattening $ {\cal M}_0 (p_1,0,0): $
   \be % f15
  E {\bf (p)} = \frac{p_1^2}{2m_1} \left(\frac{p_x^2}{p_1^2}\right) + \frac{p_2^2}{m_2}
  \left(\frac{p_y^2 + p_z^2}{p_2^2}\right)^l
   \ee
   where $p_1,p_2 $ have dimensions of  momentum. When $ l = 1$ this expression
    corresponds to the ellipsoidal FS, and $ m_1, m_2 $ are the principal values
    of the effective mass tensor. The FS curvature equals:
   \bea % f16
\!\!\!\!&&\!\! K{\bf(p)}= \frac{l}{m_2v^4} \left(\frac{p_y^2 +
p_z^2}{p_2^2}\right)^{l-1}
  \nn\\ \!\! \!\!\!& &\!\!\!\! \times\!
\left[\frac{1}{m_1}(v_y^2 + v_z^2) + v_x^2 \frac{l(2l -1)}{m_2}
\left(\frac{p_y^2 +p_z^2}{p_2^2}\right)^{l-1}\right].
    \eea
   For  $ l > 1$ the curvatures of both principal cross sections of the FS
    become zero at the point $ (p_1,0,0) $ indicating the FS local flattening.

Turning to the spherical coordinates in the velocity space we can
rewrite the expression (16) in the form:
  \be % f17
 K(\theta,\varphi) = W (\theta, \varphi) (\cos^2\theta + \sin^2 \varphi)^{(1-\beta)/2}
    \ee
      where the shape parameter $ \beta = -1 +2/(2l - 1). $ When $ l > 1, $ the
      FS curvature becomes zero at $ \theta = \pi/2,\ \varphi = 0 $ which correspond
      to the point $ {\cal M}_0 . $ The parameter $ \beta $ takes on values from the
      interval $ (-1,1) $, and the more pronounced is the FS flattening near the point
      $ {\cal M}$ ( the greater is the value of $l) $ the closer is $ \beta $ to $ -1. $
      The ``anomalous"  contribution to the conductivity originating from the flattened
      segment of the FS has the form similar to Eq. (13), namely:
 \be %f18
 \sigma_a (\omega,q) = \mu \sigma_0 (q) \left (\frac{\omega}{q
 v(\pi/2,0)}    \right)^\beta
 \left[1 -i\tan \left(\frac{\pi\beta}{2}\right)\right].
  \ee
  Here, $ \mu $ is a small dimensionless factor proportional to the relative number
  of conduction electrons associated with the flattened part of the FS. Due to the
  smallness of $ \mu $ the term (17) may be significant only when $ \beta \leq 0\
  (l \geq 1.5). $ Otherwise, it could be neglected.

Now, we proceed in calculations of the surface impedance given by the expression (2).
Under anomalous skin effect conditions the impedance can be represented as an
expansion in inverse powers of the anomaly parameter $ (\xi \gg 1). $ Representing
the conductivity as the sum of terms (5) and (6), we can calculate two first terms
in the expansion of the surface impedance in inverse powers of the anomaly parameter:
     \bea % f19,16,10,12
 Z &\equiv &  R - i H = - \frac{8 i \omega}{c^2} \delta \int_0^\infty d t
\frac{1}{1 - i t^3 (1 + i t/\xi)}  \nn \\ \nn \\ & \approx &
 Z_0 \left(\frac{\omega}{\omega_0}\right)^{2/3}
\left[1 -i\sqrt3 - \frac{2}{3} \left (\frac{\omega}{\omega_0}\right)^{2/3}
\big(1 + \sqrt 3 \big) \right ] \nn \\
           \eea
 where $ \delta = (c^2 \hbar^3/ e^2 p_0^2 \omega)^{1/3} $ is the skin depth,
  $$
  \ds Z_0 = \frac{8 \pi}{3 \sqrt 3} \frac{v_0}{c^2}; \qquad \xi =
  \frac{v_0}{\omega \delta} \equiv \left(\frac{\omega}{\omega_0}\right)^{2/3}
  $$
  is the anomaly parameter, and frequency $ \omega_0 $ equals:
  \be % f20,17,11
\omega_0 = \left (\frac{v_0}{\hbar} \right)^{3/2}
\frac{e p_0}{c} .
   \ee
 Keeping in mind that $ v_0 \sim v_F $ and $ p_0 \sim p_F $ we can roughly
 estimate the characteristic frequency $ \omega_0. $ In good metals the
 electron density has the order of $ 10^{21}-10^{22}$ cm$^{-3}$, so
 $ \omega_0 \sim 10^{12}-10^{13} s^{-1}. $ This is significantly smaller
 that the plasma frequency $ \omega_p $ which in good metals is of the
 order of $ 10^{15}-10^{16} s^{-1}. $ As one would expect, the inequality
 $ \omega \ll \omega_0 \ (\xi \gg 1) $ agrees with the general requirement
 on frequencies $ \omega \ll \omega_p, $ and could be satisfied at
 $ \omega \sim 10^{10}-10^{11} s^{-1}. $

The expression of the form (19) was first obtained by R. B. Dingle
(see \cite{3}) within the isotropic model of metal. Later it was
generalized to be applied to realistic metals, assuming that their
FSs do not include segments of zero curvature \cite{8}. For such FSs
the frequency dependence of the surface impedance has the same
character, as for a Fermi sphere. The main approximation of the
surface impedance is proportional to $ \omega^{2/3} $ and the first
correction to it is proportional to $ \omega^{4/3}. $
 \begin{figure}[t]
\begin{center}
\includegraphics[width=8.6cm,height=4.4cm]{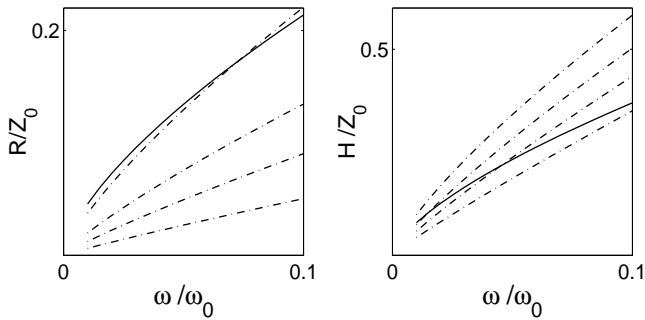}
%[width=5.0cm,height=6.6cm]{M1.eps}%{g2.eps}
\caption{Frequency dependence of the real $ ( R) $ and imaginary $
(H) $ parts of the surface impedance of a metal under the anomalous
skin effect. Dashed-dotted lines are plotted using Eq.(12) at $ \rho
= 0.2 $ and $ \beta = -0.8,-0.7,-0.6,-0.4, $ from the top to the
bottom. Solid lines represent real and imaginary parts of the
surface impedance of a metal whose FS does not include nearly
cylindrical and/or flattened segments. } \label{rateI}
\end{center}
\end{figure}

When the FS includes locally flattened or
nearly cylindrical segments the asymptotic expression for the
surface impedance changes. The effect of this anomalous
local geometry of the FS on the impedance is especially strong for
$ \beta < -0.5, \ \rho \stackrel{ >}{\sim} 0.1. $ Under these
conditions the ``anomalous" contribution dominates over the other
terms in the expression for conductivity and determines the principal
term of the surface impedance. As a result we have:
  \be %f21,18,12
 Z \approx Z_0 \zeta (\beta) \left (\frac{\omega}{\omega_0} \right)^{2/(3+ \beta)}
    \ee
 where
  \bea % f22,19,13
\zeta (\beta) &=& \frac{3 \sqrt 3 \, \rho^{-1/3 + \beta}}{3 + \beta}
\left( \cos\frac{\pi \beta}{2} \right)^{1/(3+\beta)}
   \nn\\\nn\\ &\times&
\left[ \cot \left(\frac{\pi}{3 + \beta}\right) - i \right].
    \eea

 The surface impedance described with (21) differs in magnitude from that
 of a conventional metal whose FS does not include zero curvature segments.
 Frequency dependence of the surface impedance also changes as shown in
 the Fig. 2. Now it is proportional to $\omega^{2/(\beta+3)}. $ For a
 nearly cylindrical effective strip the exponent $ 2/(\beta + 3) $ varies
 in the interval $ (0.8,1) $ where the value $1$ corresponds to the
 precisely cylindrical strip. So, when the significant part of effective
 electrons is associated with a nearly cylindrical effective strip, the
 impedance should slower increase with increase of frequency than in a
 ``conventional" case.
\begin{figure}[t]
\begin{center}
\includegraphics[width=8.6cm,height=4.4cm]{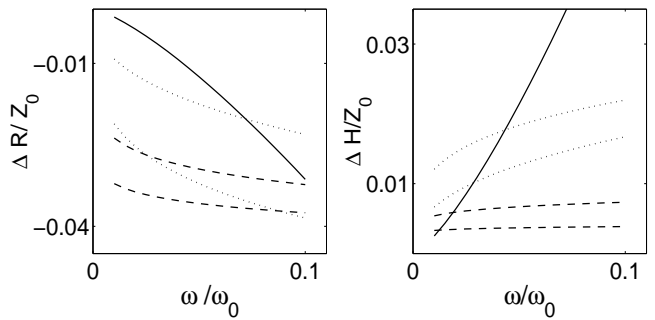}
%[width=5.0cm,height=6.6cm]{M1.eps}%{g2.eps}
\caption{ Frequency dependence of the real $ (\Delta R) $ and
imaginary $ (\Delta H) $ parts of the first correction to the main
term in the surface impedance expansion in the inverse powers of the
anomaly parameter. The curves are plotted at $ \rho= 0.01, \ \beta =
-0.9; \ \rho = 0.02, \ \beta = - 0.8 $ (dashed lines); $ \rho = 0.1,
\ \beta = -0.5; \ \rho = 0.1, \ \beta = - 0.4 $ (dotted lines).
Solid lines present the real and imaginary parts of the first
correction to the main approximation
 for the impedance of a metal whose FS does not include nearly cylindrical
 and/or flattened segments.}
 \label{rateI}
\end{center}
\end{figure}

 Now we consider more realistic case when either $ \rho $ or $ \beta $ or
 both take on values close to zero (a zero curvature segment on the effective
 part of the FS is narrow and/or the curvature anomaly is only moderately
 pronounced). In this case the anomalous contribution (13) is the first
 correction to the main approximation for the conductivity, and it
 determines the first correction to the approximation for the surface impedance:
  \be %f23,20,14
Z = Z_0 \left (\frac{\omega}{\omega_0} \right)^{2/3} (1 - i \sqrt 3) + \Delta Z.
    \ee
 Here,
  \bea %f24-25,21-22,15-16
 \Delta Z & \equiv & \Delta R - i \Delta H \approx - Z_0 \eta (\beta)
 \left (\frac{\omega}{\omega_0} \right)^{2(\beta + 1)/3};
    \\ \nn \\
\eta(\beta) & = &  \frac{\rho(\beta+1)}{\sqrt 3}
\frac{1}{\cos (\pi \beta/2)}
\bigg \{\cot \bigg( \frac{ \pi (\beta + 1)}{3} \bigg) + i \bigg \} . \nn \\
   \eea

 We can apply this result (24) to describe the contribution to the
 surface impedance from narrow or weakly developed nearly cylindrical
 strip and also from a point of flattening located on the effective segment
 of the FS. The correction to the main approximation of the surface
 impedance now is proportional to $ \omega^{2(\beta+1)/3}$ as we show
 in the Fig. 3. The presented analysis shows that the surface impedance
 of a semiinfinite metal whose FS has points or lines of zero  curvature
 can be described by the formulas (21)--(24). The obtained asymptotic
 expressions indicate that an anomaly of curvature on an effective line
 changes frequency dependence of the surface impedance, and under certain
 conditions it can essentially change its magnitude. This follows from
 the discussed above relation between the curvature of the FS and the
 number of effective electrons.
%%%%%%%%%%%%%%%%%%%%%%%%%%%%

\section{III. conclusion}

 The concept of Fermi surface is recognized as one of the most
meaningful concepts in condensed matter physics, providing  an
excellent insight in the main electronic properties of conventional
metals and other materials with metallic-like conductivity.
Extensive studies based on experimental data concerning effects
responsive to the structure of electronic spectra in metals and
using advanced computational methods were carried out to restore the
FS geometries. These efforts were resulted in the impressive mapping
of the FSs of conventional metals. However, in the course of these
studies a comparatively little attention was paid to fine local
features in the FS geometries including zero curvature lines and/or
points of flattening. These local curvature anomalies do not
significantly affect main geometrical characteristics of FSs (such
as connectivity, locations of open orbits, sizes and arrangements of
sheets) which are usually determined from the standard experiments.
Therefore these local features could be easily missed when a FS is
restored, if one does not expect them to be present, and does not
pay special attention to keep them in the resulting FS image. So, it
is important to explore possible experimental manifestations of the
FS curvature anomalies. Adoption of the phenomenological models is
justified in these studies as far as these models are based on
reasonable assumptions concerning the FS geometry. Actually,
phenomenological models were commonly used to develop the theory of
``standard" effects such as de Haas-van Alphen effect, which were
(and are) employed as tools to obtain informations concerning FSs
shapes \cite{19}. This approach to fermiology does not contradict
that one based on electron band structure calculations. It
supplements the latter. Such supplementing analysis could bring new
insight in the physical nature and origin of some physical effects
including these considered in the present work, and show their
usefulness in studies of the FSs geometries.

As for the particular models  adopted in the present work they may
be reasonably justified within a nearly free electron approximation.
Adopting the nearly free-electrons approach we arrive at the
energy-momentum relation for conduction electrons:
  \be % f26
 E = \frac{\bf k^2}{2m} + \frac{\bf g^2}{2m} - \frac{1}{m} \sqrt{{\bf
 (k\cdot g)^2} +  m^2V^2 },
  \ee
where $ m $ is the effective mass, $ {\bf k= g - p;\ g = \hbar G}/2;
\ \bf G $ is a reciprocal lattice wave vector; $ V $ is Fourier
component of the potential energy of electron in the lattice field
which corresponds to the vector $ \bf G. $
  Within the nearly free-electron model the energy $ V $ is assumed
  to be small compared  to the Fermi energy $ E_F ,$ so we introduce
  a small parameter $ \epsilon = \sqrt{V/E_F} $.

The corresponding FS looks like a sphere with ``knobs" located at
those segments which are close to the boundaries of the Brillouin
zone. Inflection lines of zero curvature pass along the boundaries
between the knobs and the main body of the FS. A FS segment
including a knob and its vicinity is axially symmetric, and the
symmetry axis is directed along the corresponding reciprocal lattice
vector. In further analysis we single out such segment to consider
it separately. For certainty we choose the coordinate system whose
$``x"$ axis is directed along the reciprocal lattice vector. Within
the chosen segment the FS curvature is described with the
expression:
    \be % f27
 K = \frac{m^2 v_x^2 + p_\perp^2 dv_x/dp_x}{(p_\perp^2 + m^2 v_x^2)^2}
  \ee
 where $ p_\perp^2 = p_y^2 + p_z^2. $

Equating the FS curvature to zero, and using the energy-momentum
relation (26) we find the values $p_{x0} $ and $p_{\perp 0} $
corresponding to the inflection line.
 We get:
   \be % f28
 p_{x0} = p_F  (1 - \epsilon^2/\sqrt 2), \qquad p_{\perp 0} \approx p_F \epsilon,
  \ee
  where $p_F $ is the radius of the original Fermi sphere. Now, we can expand
  the variable $ p_x $ in powers of $(p_\perp - p_{\perp 0}) $ near the zero
  curvature line. Taking into account that $ d^2 p_x/ dp_\perp^2 $ turns zero
  at points belonging to the inflection line, and keeping the lowest-order
  terms in the expansion, we obtain:
   \be %f29
        p_x \approx p_{x0} - \frac{\epsilon}{\sqrt 2} (p_\perp - p_{\perp 0}) -
  \frac{p_F}{\sqrt 2\, \epsilon} \left(\frac{p_\perp - p_{\perp 0}}{p_F}\right)^3.
   \ee
    Substituting this approximation into Eq. (26) we arrive at the following
    energy-momentum relation:
  \be % f30
  E{\bf (p)} = \frac{p_x^2}{2m} + \frac{2}{\epsilon} \frac{p_F^2}{2m}
  \left(\frac{p_\perp - p_{\perp 0}}{p_F} \right)^3.
   \ee
  The latter could be employed near the zero curvature line where
  $ p_{\perp 0} \ll p_F. $ Omitting $ p_{\perp 0} $ we arrive at the
  energy-momentum relation of the form (15) where $ l = 3/2 . $ Also,
  we can compare the equations describing cross-sections $ p_y = 0 $
  of the FS corresponding to the Eq. (30) and our phenomenological
  model (8), (9). Again, we see that these equations are in
  agreement with each other provided that $ l = 3/2. $

There is an experimental evidence that "necks" connecting
quasispherical pieces of the FS of copper include nearly cylindrical
belts \cite{19}. It is also likely that the FS of gold possesses the
same geometrical features for it closely resembles that of copper.
As for possible flattening of the FS, experiments of \cite{20,21,22}
on the cyclotron resonance in a magnetic field normal to the metal
surface give grounds to conjecture that such anomalies could be
found on the FSs of cadmium, zinc and even potassium.
 Another group of materials where we can expect the FS curvature anomalies
 to be manifested includes layered structures with metallic-type conductivity
 (e.g. $\alpha - (BEDT - TTF)_2 Mhg(SCH)_4 $ group of organic metals). Fermi
 surfaces of these materials are sets of rippled cylinders, isolated or
 connected by links. There exists experimental evidence that the
 quasi-two-dimensional FSs of some organic metals include segments
 with zero curvature \cite{23}.
 Also, recent investigations give grounds to expect the FSs of some new
conducting materials include flattened segments \cite{24,25,26}.

The most important result of the present work is that it shows how
such fine geometric features as points of flattening and/or zero
curvature lines could be manifested in experiments on the anomalous
skin effect. It is shown that when the FS includes nearly
cylindrical segments or it is flattened at some points, qualitative
changes may occur in frequency dependencies of the surface impedance
under the anomalous skin effect. Being observed in experiments, such
unusual frequency dependencies would indicate the presence of
zero-curvature lines and points on the FS, and display their
location. Also, analyzing these frequency dependencies, the shape
parameter $ \beta $ could be found giving additional information on
the FSs local structure. This information may be used in further
studies of the FSs geometries.
% \vspace{2mm}

\section{ Acknowledgments} The author thanks G. M. Zimbovsky for
help with the manuscript. This work was supported  by NSF Advance
program SBE-0123654, NSF-PREM 0353730, and PR Space Grant
NGTS/40091.

\end{document}